\documentclass[12pt]{article}
\usepackage{graphicx}

\newcommand\pubnumber{UdeM-GPP-TH-13-225}
\newcommand\pubdate{\today}

\def\udem{Physique des Particules, Universit\'e de
  Montr\'eal,\\ C.P. 6128, succ.\ centre-ville, Montr\'eal, QC, Canada
  H3C 3J7}
\def\cegep{D\'epartement de physique, C\'egep de Saint-Laurent,
  \\ 625, avenue Sainte-Croix, Montr\'eal, QC, Canada H4L 3X7}
\def\support{\footnote{This work was financially supported by
    NSERC of Canada (BB, DL) and by FQRNT du Qu\'ebec (MI).}}

\def\Title#1{\begin{center} {\Large #1 } \end{center}}
\def\Author#1{\begin{center}{ \sc #1} \end{center}}
\def\Address#1{\begin{center}{ \it #1} \end{center}}
\def\andauth{\begin{center}{and} \end{center}}

\newcommand\pubblock{\rightline{\begin{tabular}{l} \pubnumber\\
         \pubdate  \end{tabular}}}
\newenvironment{Abstract}{\begin{quotation}  }{\end{quotation}}
\newenvironment{Presented}{\begin{quotation} \smallskip 
      \begin{center}\begin{large}}{\end{large}\end{center} \end{quotation}}
\def\Acknowledgements{\bigskip  \bigskip \begin{center} \begin{large}
             \bf ACKNOWLEDGEMENTS \end{large}\end{center}}

\def\beq{\begin{equation}}
\def\eeq#1{\label{#1}\end{equation}}
\def\eeqn{\end{equation}}

\def\beqa{\begin{eqnarray}}
\def\eeqa#1{\label{#1}\end{eqnarray}}
\def\eeqan{\end{eqnarray}}

\let\bar=\overbar

\def\Dslash{\not{\hbox{\kern-4pt $D$}}}
\def\dslash{\not{\hbox{\kern-2pt $\del$}}}

\def\msb{{\bar{\ssstyle M \kern -1pt S}}}

\begin{document}
\begin{titlepage}
\pubblock

\vfill
\Title{Measurement of $\gamma$ using $B \to K \pi \pi$ and $B \to KK{\bar K}$ decays}
\vfill
\Author{Bhubanjyoti Bhattacharya, David London}
\Address{\udem}
\andauth
\Author{Maxime Imbeault\support}
\Address{\cegep}
\vfill
\begin{Abstract}
The {B{\sc a}B{\sc ar}} measurements of the Dalitz plots for $B^0 \to
K^+\pi^0\pi^-$, $B^0 \to K^0\pi^+\pi^-$, $B^+ \to K^+\pi^+\pi^-$, $B^0
\to K^+ K^0 K^-$, and $B^0 \to K^0 K^0 {\bar K}^0$ decays are used to
cleanly extract the weak phase $\gamma$. We find four possible
solutions: $(31^{+2}_{-3})^\circ$, $(77 \pm 3)^\circ$,
$(258^{+4}_{-3})^\circ$, and $(315^{+3}_{-2})^\circ$. One solution --
$(77 \pm 3)^\circ$ -- is consistent with the SM. Its error, which
includes leading-order flavor-SU(3) breaking, is far smaller than that
obtained using two-body $B$ decays.
\end{Abstract}
\vfill
\begin{Presented}
Talk given by David London at the 2013 Flavor Physics and CP Violation
conference (FPCP-2013), Buzios, Rio de Janeiro, Brazil, May 19-24,
2013. \\ Talk based on arXiv:1303.0846 \cite{3bodyPRL}.
\end{Presented}
\vfill
\end{titlepage}
\def\thefootnote{\fnsymbol{footnote}}
\setcounter{footnote}{0}

The standard way to obtain clean information about CKM phases is
through the measurement of indirect CP violation in $B^0/{\bar B}^0
\to f$. The conventional wisdom is that one cannot obtain such clean
information from 3-body decays.  There are two reasons. First, $f$
must be a CP eigenstate. While this holds for certain 2-body final
states (e.g., $J/\psi K_S$, $\pi^+\pi^-$, etc.), 3-body states are, in
general, not CP eigenstates. For example, consider $K_S \pi^+\pi^-$:
the value of its CP depends on whether the relative $\pi^+\pi^-$
angular momentum is even (CP $+$) or odd (CP $-$). Second, one can
only cleanly extract a weak phase using indirect CP asymmetries if the
decay is dominated by amplitudes with a single weak phase. But 3-body
decays generally receive significant contributions from amplitudes
with different weak phases. Even if the CP of the 3-body final state
could be fixed in some way, we would still need a way of dealing with
this ``pollution.''

Recently it was shown that all of these difficulties can be overcome
\cite{3body1,3body2,3body3}. There are three ingredients. First, the
Dalitz plots of the 3-body decays are used to separate CP $+$ and CP
$-$ final states. Second, the decay amplitudes are expressed in terms
of diagrams. This permits the removal of the above pollution. Third,
the electroweak-penguin (EWP) and tree diagrams are related, which
reduces the number of unknown parameters. These three points are
discussed below.

In the decay $B \to P_1 P_2 P_3$, one defines the three Mandelstam
variables $s_{ij} \equiv \left( p_i + p_j \right)^2$, where $p_i$ is
the momentum of $P_i$.  (The three $s_{ij}$ are not independent, but
obey $s_{12} + s_{13} + s_{23} = m_B^2 + m_1^2 + m_2^2 + m_3^2$.) The
Dalitz plot is given in terms of two Mandelstam variables, say
$s_{12}$ and $s_{13}$. The key point is that, using the Dalitz plot,
one can reconstruct the full decay amplitude ${\cal M}(B \to P_1 P_2
P_3)(s_{12},s_{13})$.

The amplitude for a state with a given symmetry is then found by
applying this symmetry to ${\cal M}(s_{12},s_{13})$. For example, the
amplitude for the final state $K_S \pi^+ \pi^-$ with CP $+$ is
symmetric in $2 \leftrightarrow 3$. This is given by $[{\cal
    M}(s_{12},s_{13}) + {\cal M}(s_{13},s_{12})]/\sqrt{2}$. This
amplitude is then used to compute all the observables for the
decay. Note: all observables are momentum dependent -- they take
different values at each point in the Dalitz plot.

\begin{figure}[!htbp]
	\centering
		\includegraphics[height=3.5cm]{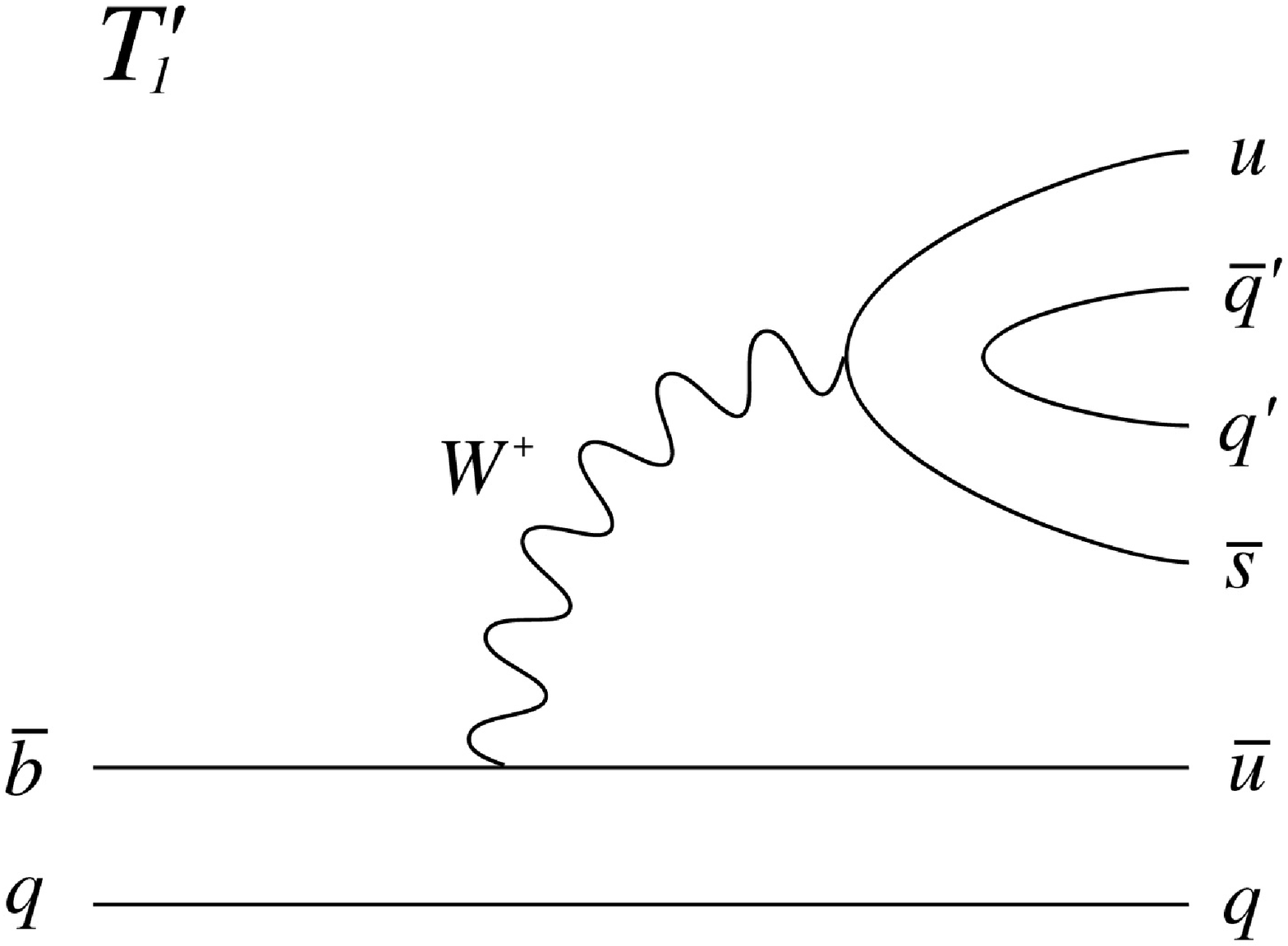}
		~~~~~~~ \includegraphics[height=3.5cm]{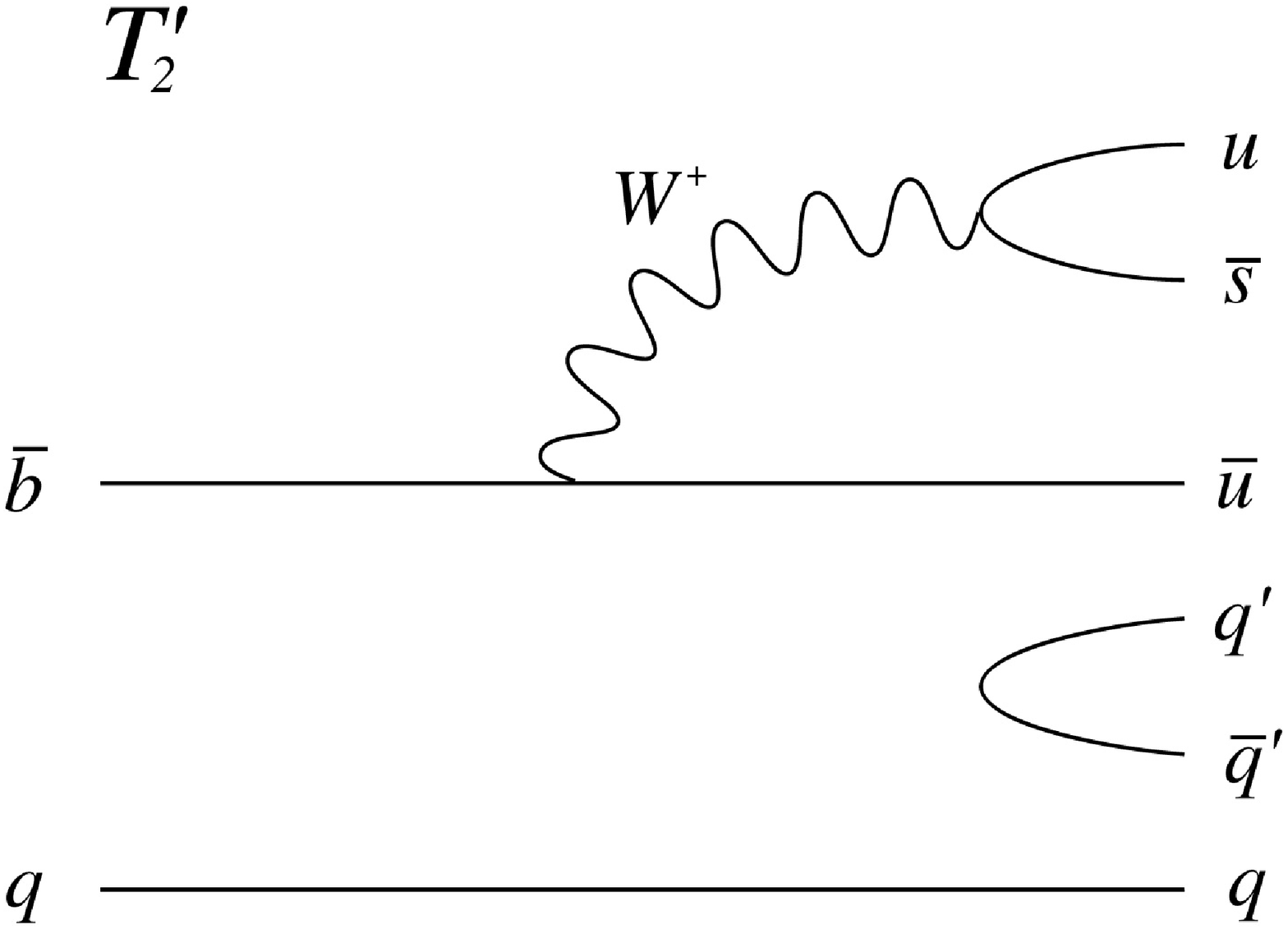}
\caption{$T'_1$ and $T'_2$ diagrams contributing to $B \to K \pi
  \pi$.}
\label{diagrams}
\end{figure}

In order to remove the pollution due to additional decay amplitudes,
one first expresses the full amplitude in terms of diagrams
\cite{3body1}. These are similar to those of two-body $B$ decays ($T$,
$C$, etc.), but here one has to ``pop'' a quark pair from the vacuum.
We add the subscript ``1'' (``2'') if the popped quark pair is between
two non-spectator final-state quarks (two final-state quarks including
the spectator). Fig.~\ref{diagrams} shows the $T'_1$ and $T'_2$
diagrams contributing to $B \to K \pi \pi$ (as this is a ${\bar b} \to
{\bar s}$ transition, the diagrams are written with primes).

Note: unlike the 2-body diagrams, the 3-body diagrams are momentum
dependent. This must be taken into account whenever the diagrams are
used.

In $B \to K \pi$ decays, under flavor SU(3) symmetry there are
relations between the EWP and tree diagrams \cite{NRGPY}. In
Ref.~\cite{3body2} it was shown that similar EWP-tree relations hold
for $B \to K \pi \pi$ decays. Taking $c_1/c_2 = c_9/c_{10}$ for the
Wilson coefficients (which holds to about 5\%), these take the simple
form (the exact relations are given in Ref.~\cite{3body2})
\begin{equation}
P'_{EWi} = \kappa T'_i ~~,~~~~ P^{\prime C}_{EWi} = \kappa C'_i ~~~~(i=1,2) ~, 
\end{equation}
where
\begin{equation}
\kappa \equiv - \frac{3}{2} \frac{|\lambda_t^{(s)}|}{|\lambda_u^{(s)}|} \frac{c_9+c_{10}}{c_1+c_2} ~,
\end{equation}
with $\lambda_p^{(s)}=V^*_{pb} V_{ps}$.

However, there is an important caveat. Under SU(3), the final state in
$B \to K \pi \pi$ involves three identical particles, so that the six
permutations of these particles must be taken into account.  But the
EWP-tree relations hold only for the totally symmetric state. This
state, ${\cal M}_{\rm fs}$ (`fs' = `fully symmetric'), is found by
symmetrizing ${\cal M}(s_{12},s_{13})$ under all permutations of
1,2,3.  The analysis must therefore be carried out for this state.

With the above three ingredients, one can cleanly extract weak-phase
information from 3-body decays. The fundamental idea is as follows.
It is common to combine observables from different 2-body $B$ decays
in order to extract weak-phase information. Examples include obtaining
$\alpha$ from $B\to\pi\pi$ \cite{GL}, obtaining $\gamma$ from $B\to
DK$ \cite{GLW,ADS,GGSZ}, and observing the ``$B\to\pi K$ puzzle'' in
$B\to\pi K$ \cite{BpiK}.  In 3-body $B$ decays, the idea is the same,
except that the analysis applies to each point in the Dalitz
plot. (That is, the analysis is momentum dependent.) The disadvantage
is that the analysis is more complicated. However, there is a big
advantage: since it holds at each point in the Dalitz plot, the
analysis really represents many independent determinations of the
weak-phase information. These can be combined, considerably reducing
the error. Below we present an example of such an analysis involving
$B \to K \pi \pi$ and $B \to KK{\bar K}$ decays \cite{3body3}.

We consider the five decays $B^0 \to K^+\pi^0\pi^-$, $B^0 \to
K^0\pi^+\pi^-$, $B^+ \to K^+\pi^+\pi^-$, $B^0 \to K^+ K^0 K^-$, and
$B^0 \to K^0 K^0 {{\bar K}^0}$. The $B \to K \pi \pi$ amplitudes are written in
terms of diagrams with a popped $u{\bar u}$ or $d{\bar d}$ quark pair
(these are equal under isospin), while the diagrams of the $B \to KK{\bar K}$
amplitudes have a popped $s{\bar s}$ pair. But flavor-SU(3) symmetry,
which is needed for the EWP-relations, implies that all diagrams are
equal. All five amplitudes are therefore written in terms of the same
diagrams.

Note, however, that flavor-SU(3) symmetry is not exact.  It is
therefore important to keep track of a possible difference between
$B \to K \pi \pi$ and $B \to KK{\bar K}$ decays.

The expressions for the amplitudes in terms of diagrams are given in
Ref.~\cite{3body3}. The diagrams can be combined into ``effective
diagrams'' \cite{3bodyPRL}:
\begin{eqnarray}
a \equiv - {\tilde P}'_{tc} + \kappa \left(\frac23 T'_1 + \frac13 C'_1
+ \frac13 C'_2 \right) ~,~~~ \nonumber\\
b \equiv T'_1 + C'_2 ~,~~
c \equiv T'_2 + C'_1 ~,~~
d \equiv T'_1 + C'_1 ~.
\label{effdiag}
\end{eqnarray}
The decay amplitudes can now be written in terms of five diagrams,
$a$-$d$ and ${\tilde P}'_{uc}$:
\begin{eqnarray}
\label{effamps}
2 A(B^0 \to K^+\pi^0\pi^-)_{\rm fs} &=& b e^{i\gamma} - \kappa c ~, \nonumber\\
\sqrt{2} A(B^0 \to K^0\pi^+\pi^-)_{\rm fs} &=& -d e^{i\gamma} - {\tilde P}'_{uc} e^{i\gamma} - a + \kappa d ~, \nonumber\\
\sqrt{2} A(B^+ \to K^+ \pi^+ \pi^-)_{\rm fs} &=& -c e^{i\gamma} -{\tilde P}'_{uc} e^{i\gamma} - a + \kappa b ~, \nonumber\\
\sqrt{2} A(B^0 \to K^+ K^0 K^-)_{\rm fs} &=& \alpha_{SU(3)} (-c e^{i\gamma} -{\tilde P}'_{uc} e^{i\gamma} - a + \kappa b ) ~, \nonumber\\
A(B^0 \to K^0 K^0 {{\bar K}^0})_{\rm fs} &=& \alpha_{SU(3)} ({\tilde P}'_{uc} e^{i\gamma} + a ) ~.
\end{eqnarray}
In the above, $\alpha_{SU(3)}$ measures the amount of flavor-SU(3)
breaking between $B \to K \pi \pi$ and $B \to KK{\bar K}$ decays,
i.e., between diagrams with a final-state $u{\bar u}$/$d{\bar d}$
quark pair and those with an $s{\bar s}$ pair. It must be stressed
that $\alpha_{SU(3)}$ is only a leading-order SU(3)-breaking term. For
example, it assumes that the SU(3) breaking is the same for all
diagrams. The possible effect of next-to-leading-order SU(3) breaking
must be kept in mind.

Now, in the flavor-SU(3) limit, $\alpha_{SU(3)} = 1$ (the imaginary
piece vanishes in this limit), so that we have $A(B^+ \to K^+ \pi^+
\pi^-)_{\rm fs} = A(B^0 \to K^+ K^0 K^-)_{\rm fs}$. This implies that
the $B^+$ decay does not furnish any new information. The remaining
four amplitudes depend on 10 theoretical parameters: 5 magnitudes of
diagrams, 4 relative phases, and $\gamma$.  But there are 11
experimental observables: the decay rates and direct asymmetries of
each of the 4 processes, and the indirect asymmetries of $B^0 \to
K^0\pi^+\pi^-$, $B^0 \to K^+ K^0 K^-$ and $B^0 \to K^0 K^0 {{\bar
    K}^0}$.  With more observables than theoretical parameters,
$\gamma$ can be extracted from a fit.

If one allows for SU(3) breaking ($|\alpha_{SU(3)}| \ne 1$), we can
add two more observables: the decay rate and direct CP asymmetry for
the $B^+$ decay. In this case it is possible to extract $\gamma$ even
with the inclusion of $|\alpha_{SU(3)}|$ as a fit parameter. (Note
that the observables are insensitive to the phase of
$\alpha_{SU(3)}$.)

Since the diagrams and observables are all momentum dependent, this
implies that the above method for extracting $\gamma$ in fact applies
to each point in the Dalitz plot. However, since the value of $\gamma$
is independent of momentum, the method really represents {\it many}
independent measurements of $\gamma$. These can be combined, reducing
the error on $\gamma$.

The observables are obtained as follows. The $B \to P_1 P_2 P_3$
amplitude is written as
\begin{equation}
{\cal M} (s_{12}, s_{13}) = {\cal N}_{\rm DP}\sum\limits_j c_j e^{i\theta_j} F_j
(s_{12}, s_{13})~,
\end{equation}
where the index $j$ runs over all resonant and non-resonant
contributions. Each contribution is expressed in terms of isobar
coefficients $c_j$ (amplitude) and $\theta_j$ (phase), and a dynamical
wave function $F_j$. The $F_j$ take different forms depending on the
contribution. The $c_j$ and $\theta_j$ are extracted from a fit to the
Dalitz-plot event distribution.

{B{\sc a}B{\sc ar}} has performed such fits for the five decays of
interest \cite{Expt}. For each decay, given the $c_j$, $\theta_j$ and
$F_j$, we reconstruct the amplitude for that decay as a function of
$s_{12}$ and $s_{13}$. We then obtain ${\cal M}_{\rm fs}$ by
symmetrizing under all permutations of 1,2,3. This process is repeated
for the CP-conjugate process, where we construct $\overline{{\cal
    M}}_{\rm fs}$.

The experimental observables are then obtained as follows:
\begin{eqnarray} \label{XYZdef}
X(s_{12}, s_{13}) &=& |{\cal M}_{\rm fs}(s_{12}, s_{13})|^2 + |\overline
{{\cal M}}_{\rm fs}(s_{12}, s_{13})|^2~, \nonumber \\
Y(s_{12}, s_{13}) &=& |{\cal M}_{\rm fs}(s_{12}, s_{13})|^2 - |\overline
{{\cal M}}_{\rm fs}(s_{12}, s_{13})|^2~, \nonumber \\
Z(s_{12}, s_{13}) &=& {\rm Im}\left[{\cal M}^*_{\rm fs}(s_{12}, s_{13})
~\overline{{\cal M}}_{\rm fs}(s_{12}, s_{13})\right]~.
\end{eqnarray}
The experimental error bars on these quantities are found by varying
the input isobar coefficients over their $1\sigma$-allowed ranges.
The effective CP-averaged branching ratio ($X$), direct CP asymmetry
($Y$), and indirect CP asymmetry ($Z$) may be constructed for every
point on any Dalitz plot.  However, $Z$ can be measured only for $B^0$
decays to a CP eigenstate.

There is one technical point: in its $K_SK_SK_S$ analysis, due to
limited statistics {B{\sc a}B{\sc ar}} takes $A(B^0 \to K_S K_S K_S) =
A({\bar B}^0 \to K_S K_S K_S)$. This implies that (i) $Y$ and $Z$
vanish for every point of the Dalitz plot, and (ii) the (small)
diagram ${\tilde P}'_{uc}$ must be set to zero.  The removal of an
equal number of unknown parameters (amplitude and phase of ${\tilde
  P}'_{uc}$) and observables does not affect the viability of the
method.

\begin{figure}[!htbp]
	\centering
		\includegraphics[height=5.0cm]{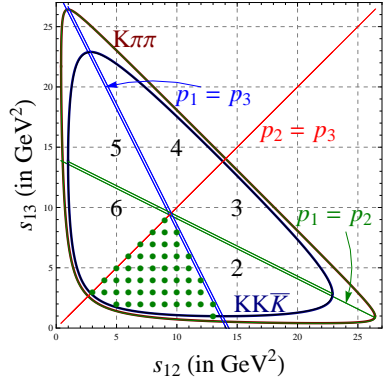}
\caption{Kinematic boundaries and symmetry axes of the $B \to K \pi
  \pi$ and $B \to KK{\bar K}$ Dalitz plots. The symmetry axes divide
  each plot into six zones.}
\label{DP}
\end{figure}

Since the amplitudes used to construct the observables are fully
symmetric under the interchange of the three Mandelstam variables,
only one sixth of the Dalitz plot provides independent information. In
order to avoid multiple counting, we divide each Dalitz plot into six
zones by its three axes of symmetry, and use information only from one
zone. This is illustrated in Fig.~\ref{DP}, which shows the kinematic
boundaries and symmetry axes of the $B \to K \pi \pi$ and $B \to
KK{\bar K}$ Dalitz plots. The 50 points in the region of overlap of
the first of six zones from all Dalitz plots are used for the $\gamma$
measurement.

We now perform a maximum likelihood analysis for extracting
$\gamma$. For each of the 50 points in the first Dalitz-plot zone, we
construct the $-2 \Delta \ln {\rm L}(\gamma)$ function, where ${\rm
  L}$ represents the likelihood. The sum of such functions over all
fifty points gives us a joint likelihood distribution. The local
minima of this function are then identified as the most-likely values
of $\gamma$. In order to find the $1\sigma$ error bar on $\gamma$ we
first shift the likelihood function along the vertical axis so that
the zero of the function corresponds to a local minimum. We then look
for the range of $\gamma$ that results in a unit shift along the
vertical axis of the $-2 \Delta \ln {\rm L}(\gamma)$ vs $\gamma$ plot.
The $1\sigma$ error bars on $\gamma$ are given by the condition that
$\Delta(-2 \Delta \ln {\rm L}(\gamma)) = 1$.

We perform 3 types of fit:
\begin{enumerate}

\item We assume that flavor SU(3) is a good symmetry, so that
  $\alpha_{SU(3)} = 1$. The fit involves only the four $B^0$ decay
  channels.

\item SU(3) breaking is allowed and treated as follows.  The ratio of
  $X$'s is constructed point by point from the Dalitz plots for
  $B^+\to K^+\pi^+\pi^-$ and $B^0 \to K^+ K^0 K^-$, giving
  $|\alpha_{SU(3)}|^2(s_{12}, s_{13})$. We use $|\alpha_{SU(3)}|$
  found in this way to correct the observables from the $B \to KK{\bar K}$
  Dalitz plots and use the corrected numbers in a new
  maximum-likelihood analysis for finding $\gamma$.

\item We consider observables from all five Dalitz plots but now
  include $|\alpha_{SU(3)}|$ as an additional unknown hadronic
  parameter.

\end{enumerate}

\begin{figure}[!htbp]
	\centering
		\includegraphics[height=5.0cm]{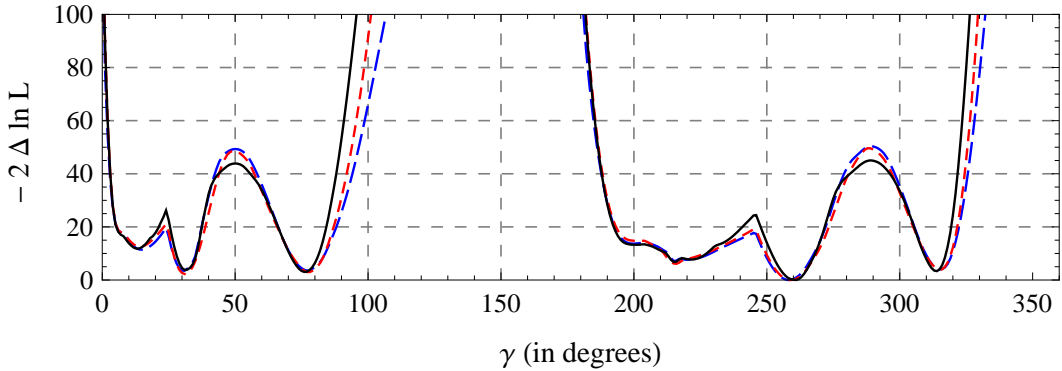}
\caption{Results of maximum-likelihood fits. The solid (black) curve
represents the fit assuming flavor-SU(3) symmetry. The short dashes (red)
represent the fit where flavor-SU(3) breaking is fixed by a point-by-point
comparison of Dalitz plots for $B^+\to K^+\pi^+\pi^-$ and $B^0\to K^+ K^0
K^-$. The long dashes (blue) represent the fit with inputs from five Dalitz
plots and an extra hadronic fit parameter $|\alpha_{SU(3)}|$.} 
\label{maxlik}
\end{figure}

The results of the maximum-likelihood fits are shown in
Fig.~\ref{maxlik}.  From this figure, we see that there is very little
difference among the three fits. This shows that, on average, SU(3)
breaking is small. This is consistent with the result from fit 2:
averaged over the 50 points, we find that $|\alpha_{SU(3)}| = 0.97 \pm
0.05$ (recall that $\alpha_{SU(3)} = 1$ corresponds to no SU(3)
breaking).

There are four preferred values for $\gamma$:
\begin{equation}
(31^{+2}_{-3})^\circ ~~,~~~~ (77 \pm 3)^\circ ~~,~~~~ (258^{+4}_{-3})^\circ ~~,~~~~ (315^{+3}_{-2})^\circ ~.
\end{equation}
Three of these indicate new physics (is this a ``$K \pi \pi$-$KK{\bar
  K}$ puzzle''?), but one solution -- $(77 \pm 3)^\circ$ -- is
consistent with the standard model.

In all cases, the error is small, 2-4$^\circ$. This can be understood
as follows.  The key point is that this method really involves 50
independent measurements of $\gamma$. Roughly speaking, if each
measurement has an error of $\pm 20^\circ$, which is somewhat larger
than other methods, then when we take a naive average, we divide the
error by $\sqrt{50}$, giving a final error of $\sim 3^\circ$.

There are several potential sources of error that have not been
included in our method. The first is correlations. Although the
errors on the isobar coefficients extracted from a given Dalitz plot
are in general correlated, such information is not always publicly
available. In our analysis we have considered the errors to be
completely uncorrelated, but we hope that a future analysis by an
experimental collaboration will take such effects into account.

Second there are possible effects from higher-order flavor-SU(3)
breaking. Such breaking may arise due to the nonzero mass difference
between pions and kaons, and between intermediate resonances. Indeed,
after the talk, Yuval Grossman expressed some skepticism about having
only one SU(3)-breaking parameter, and asked if it were possible to
include more.  Unfortunately, this cannot be done. In the method,
there are 11 observables and 9 unknown parameters (these include
$|\alpha_{SU(3)}|$). If a second SU(3)-breaking parameter were added,
there would then be 11 unknowns (these include the two magnitudes and
the relative phase of the SU(3)-breaking parameters). In this case,
with an equal number of observables and unknown parameters, one could
still extract $\gamma$, but only with even more discrete ambiguities.

This said, the error due to leading-order SU(3) breaking is small, and
so it is unlikely that the error due to higher-order SU(3) breaking is
larger. We can get a bit of a handle on this as follows. As mentioned
above, in fit 2 one obtains $|\alpha_{SU(3)}|^2(s_{12}, s_{13})$ by
computing, point by point, the ratio of $X$'s in the $B^+\to
K^+\pi^+\pi^-$ and $B^0 \to K^+ K^0 K^-$ Dalitz plots. These are then
averaged over all 50 points, giving the average value of leading-order
SU(3) breaking $|\alpha_{SU(3)}| = 0.97 \pm 0.05$. In fact, this can
be done in two different ways. One can compare the $|{\cal M}_{\rm
  fs}(s_{12}, s_{13})|^2 = (X+Y)/2$ of the two Dalitz plots to obtain
(when averaged) $|\alpha_{SU(3)}^{\cal M}|$. Alternatively, one can
compare the $|\overline {{\cal M}}_{\rm fs}(s_{12}, s_{13})|^2 =
(X-Y)/2$, giving $|\alpha_{SU(3)}^{\overline{\cal M}}|$. To leading
order, we expect $|\alpha_{SU(3)}^{\cal M}| =
|\alpha_{SU(3)}^{\overline{\cal M}}|$, so that their difference
indicates the size of higher-order SU(3) breaking. We find
$|\alpha_{SU(3)}^{\cal M}| = 0.97 \pm 0.04$ and
$|\alpha_{SU(3)}^{\overline{\cal M}}| = 0.99 \pm 0.04$, yielding a
difference of $0.02 \pm 0.06$. Though not a proof that higher-order
SU(3) breaking is small, the smallness of this difference does suggest
this conclusion.

Finally, there is one very important caveat, related to an error that
has not been included, and that can significantly affect our result.
All errors considered so far have been entirely statistical (even
SU(3) breaking). But there is also the systematic, model-dependent
error associated with the isobar analysis. This cannot be treated
statistically, i.e., reduced by averaging. This error was not given in
the {B{\sc a}B{\sc ar}} papers and so we could not include it.
Hopefully, the experimentalists themselves will redo this analysis,
including all errors.

Recall that the standard way to directly probe $\gamma$ is via $B^\pm
\to DK^\pm$ decays \cite{GLW,ADS,GGSZ}.  Although the two-body method
is expected to be theoretically clean, it is difficult experimentally,
so that the present direct measurement has a large error: $\gamma =
(66 \pm 12)^\circ$ \cite{CKMfitter}. The statistical error of
2-4$^\circ$ in the three-body method is far smaller than the two-body
error. If the systematic error is not too large, the three-body method
could well be the best way to measure $\gamma$.

To summarize: about 2-3 years ago, it was shown that, theoretically,
it is possible to cleanly extract weak-phase information from 3-body
$B$ decays. In the present study, we demonstrate that this is, in
fact, true. Using real data from {B{\sc a}B{\sc ar}}, we extract the
phase $\gamma$ from $B \to K \pi \pi$ and $B \to KK{\bar K}$
decays. We find that there is a fourfold discrete ambiguity, giving
the preferred values $\gamma = 31^\circ$, $77^\circ$, $258^\circ$ or
$315^\circ$. However, in all cases, the error is small, 2-4$^\circ$,
and includes leading-order SU(3) breaking. This is due to the fact
that, in this method, there are actually 50 independent measurements
of $\gamma$. When these are combined, the error is considerably
reduced.

The main thing that is missing is the systematic, model-dependent
error related to the isobar Dalitz-plot analysis. It is only the
experimentalists themselves who can properly include it. If the
systematic error is not too large, then this 3-body method will likely
be the best one for measuring $\gamma$.

\newpage
\Acknowledgements 
A special thank you goes to E. Ben-Haim for his important input to
this project. We also thank J. Charles, M. Gronau, N. Rey-Le Lorier,
J. Rosner, J. Smith, Y. Grossman and A. Soffer for helpful
communications. BB would like to thank G. Bell and WG IV of CKM 2012.

\end{document}